\documentclass[prl,floatfix,showpacs,amsmath,twocolumn]{revtex4}
\usepackage{amssymb}
\usepackage{graphics}
\usepackage{epsfig}
\def\qed{\leavevmode\unskip\penalty9999 \hbox{}\nobreak\hfill
     \quad\hbox{\leavevmode  \hbox to.77778em{%
               \hfil\vrule   \vbox to.675em%
               {\hrule width.6em\vfil\hrule}\vrule\hfil}}
     \par\vskip3pt}
\def\ibb #1{\leavevmode\hbox{\kern.3em\vrule
     height 1.5ex depth -.1ex width .4pt\kern-.3em\rm#1}}

\newcommand{\be}{\begin{equation}}
\newcommand{\ee}{\end{equation}}
\newcommand{\bea}{\begin{eqnarray}}
\newcommand{\eea}{\end{eqnarray}}

\newtheorem{theorem}{Theorem}

\newcommand{\SSR}{\mathrm{SSR}}

\begin{document}


\title{Nonlocal resources in the presence of Superselection Rules}
\author{N. \surname{Schuch}}
\affiliation{Max-Planck-Institut f\"ur Quantenoptik,
  Hans-Kopfermann-Str.\ 1, Garching, D-85748, Germany.}
\author{F. \surname{Verstraete}}
\affiliation{Max-Planck-Institut f\"ur Quantenoptik,
  Hans-Kopfermann-Str.\ 1, Garching, D-85748, Germany.}
\author{J.I. \surname{Cirac}}
\affiliation{Max-Planck-Institut f\"ur Quantenoptik,
  Hans-Kopfermann-Str.\ 1, Garching, D-85748, Germany.}

\begin{abstract}
Superselection rules severely alter the possible operations that
can be implemented on a distributed quantum system. Whereas the
restriction to local operations imposed by a bipartite setting
gives rise to the notion of entanglement as a nonlocal resource,
the superselection rule associated with particle number
conservation leads to a new resource, the \emph{ superselection
induced variance} of local particle number. We show that, in the
case of pure quantum states, one can quantify the nonlocal
properties by only two additive measures, and that all states with
the same measures can be asymptotically interconverted into each
other by local operations and classical communication. Furthermore
we discuss how superselection rules affect the concepts of
majorization, teleportation and mixed state entanglement.
\end{abstract}

\maketitle

One of the most remarkable and deepest discoveries in the field of
quantum information theory has been the fact that the amount of
nonlocality (i.e. entanglement) of pure bipartite quantum states
can be quantified using a single number, namely the entropy of
entanglement (EoE) \cite{BBP96a}. This is due to the fact that all
pure states with the same amount of EoE can be asymptotically
interconverted into each other by local operations and classical
communication (LOCC) and thus into singlets. The modern look at
entanglement is to see it as a \emph{resource} that allows to do
tasks that are otherwise impossible, e.g. teleportation
\cite{BBC93}, dense coding \cite{BW92} and quantum computing
\cite{Sho94}, and hence the EoE quantifies the nonlocal resources
associated to a given state.

The whole field of entanglement theory can be formally seen as the
study of the possible conversions of multipartite states given the
natural restriction that one can only implement local operations
and communicate by a classical channel (LOCC). It is evident that
the very presence of a constraint gives rise to the concept of a
resource that enables one to overcome this constraint: the essence
of e.g. teleportation is the fact that the presence of
entanglement allows to implement global quantum operations by
local means, and hence to overcome locality constraints.
Additional natural constraints should therefore give rise to new
resources and to new interesting physics and applications.

It was noted by Popescu \cite{Popescu} that such an additional
restriction applies in many physical systems in the form of a
superselection rule (SSR) \cite{WWW52}. In \cite{VC03}, we
considered the constraint of particle number conservation, which
was motivated by current quantum optical experiments on e.g. cold
atomic gases \cite{BEC}. We indeed observed that the extra
limitation leads to new applications that are impossible without
SSR: perfect data hiding \cite{TDL01} becomes possible in the
presence of SSR \cite{VC03}. On the other hand, it was shown in
\cite{VC03,KMP03} that the extra resource of a common reference
frame, in the form of a nonlocal state, allows to overcome the
restrictions imposed by the SSR (note that \cite{KMP03} also
addresses nonabelian SSR). The central contribution of this paper
is to quantify this new resource induced by SSR. Fortunately, it
turns out that there is a unique way of achieving this, and this
measure will be called \emph{superselection induced variance}
(SiV). Of course the quantification of entanglement has also to be
reconsidered in the presence of SSR (see e.g.
\cite{RS01,VC03,WV03}), but we will see that the concept of EoE
does not have to be altered. Furthermore, we will show that the
nonlocal properties of pure bipartite quantum states subject to
SSR are completely captured by specifying two (additive) measures,
namely the entropy of entanglement (EoE) and the superselection
induced variance (SiV). This follows from the fact that,
asymptotically, the Hilbert space decouples into two complementary
Hilbert spaces, each one associated to one resource. In analogy
with teleportation, the additional resource allows to implement
global quantum operations locally \cite{VC03}, although not with
unit fidelity; this defect follows from the fact that the Hilbert
space associated to the SiV has a direct sum structure as opposed
to a tensor product structure.

Let us first define the SSR under consideration more precisely.
Following \cite{VC03}, we consider a set of particles and the
corresponding Hilbert space $H$. We can always decompose
$H=\bigoplus_{N=0}^\infty H_N$ where $H_N$ is a subspace with a
total number $N$ of particles. We assume that the particle number
is a superselection observable, in the sense that the
corresponding operator $\hat{N}$ commutes with all observables.
This immediately imposes that pure states have only support in one
superselection sector $H_N$. If bipartite systems are considered
however, $H_N$ can be decomposed as $H_N=\bigoplus_{n=0}^N
(H^A_n\otimes H^B_{N-n})$, where $H^A_n$ ($H^B_n$) denotes a
Hilbert space corresponding to system A (B), with $n$ particles.
The limitation induced by the SSR precisely consists in the fact
that all local observables have to commute with the local particle
number operator. Summarized, the particle number operator is given
by $\hat N_{AB}=\hat N_{A}\otimes\openone_B+\openone_A\otimes\hat
N_B$, all physical density operators have to commute with $\hat
N_{AB}$, and all local observables with $\hat N_A$ or $\hat N_B$.
Clearly, local variations of particle number can occur as long as
the global particle number remains constant. As shown in
\cite{VC03}, the coherence between the parts of the wavefunction
with different local particle number cannot be observed locally,
and this is precisely the new kind of nonlocality that arises as a
consequence of SSR.

Let us start by considering the analogue of the Schmidt
decomposition in the presence of SSR, parameterizing all nonlocal
properties of a quantum state. It is readily verified that the
usual Schmidt decomposition on a state subject to SSR yields local
bases with all basis vectors having a fixed particle number.
Therefore we can parameterize the Schmidt coefficients as
$\{\lambda_i^n\}\equiv\{\vec{\lambda}^n\}$, where $n$ denotes
Alice's local particle number and $i$ the $i$'th Schmidt
coefficient in the corresponding subspace. When no SSR apply, the
possible conversions of one state into another one can elegantly
be described by applying the concepts of majorization
\cite{Nie99,JP99a} to the Schmidt coefficients. Let us now prove
that a similar theorem applies in the presence of SSR: given a
state $\psi$ with Schmidt coefficients $\{\vec{\lambda}^n\}$, then
this state can be converted into the set of states
$\{\phi_\alpha\}$ with corresponding probabilities $\{p_\alpha\}$
and Schmidt coefficients $\vec{\mu}_\alpha^n$ if and only if
$\forall n: \vec{\lambda}^n\prec\sum_\alpha
p_\alpha\vec{\mu}_\alpha^n$. The necessity of the condition is
obvious from the results without SSR, because all POVM elements
that one can implement have to commute with the local particle
number operator and are hence block-diagonal. Let us prove the
sufficiency for the case where there is only one state
$\phi=\phi_1$, and hence $p_1=1$. From \cite{Nie99}, it follows
that for each subspace of constant local particle number, there
exists an appropriate POVM with elements $\{M_i^n\}$, each
producing the desired state with probability $p_i^n$. The complete
conversion can therefore be obtained by taking a direct sum of the
corresponding operators: $\bar{M}_i=\bigoplus_n M_i^n$. The
problem however is that it will typically not hold that
$p_i^n=p_i^m, n\neq m$, but this defect can readily be cured by
subdividing each element $M_i^n$ into copies of itself but with
different weights such that $\forall n,m:  p_i^n=p_i^m$, which
ends the proof. The more general proof follows immediately.

Let us next try to formulate an asymptotic version of the previous
theorem. We know that, in the case of an asymptotic amount of
copies of a given state and no SSR, the majorization criterion
\emph{converges} to the entropic criterion \cite{BBP96a}.
Analogously, consider the probability distribution associated to
the variation in local particle number $(p_n=\sum_i\lambda_i^n)$.
In the case of an asymptotic amount of copies, the central limit
theorem dictates that this distribution becomes Gaussian and can
completely be characterized by its mean and variance. Let us
therefore define the measure \emph{superselection induced
variance} (SiV) for a pure bipartite state as the variance in the
local particle number:
\[V(\psi)=4\left(\langle\psi|\hat N_A^2\otimes \openone_B|\psi\rangle-\langle\psi|\hat N_A\otimes \openone_B|\psi\rangle^2\right)\]
The factor was included to normalize the SiV such that
$V(|01\rangle+|10\rangle)=1$. Note that the SiV is additive
[$V(\psi\otimes\phi)=V(\psi)+V(\phi)$] and symmetric under
interchange of $A$ and $B$. It can also readily be proven that the
SiV is an entanglement monotone \cite{Vid00}, since after any
local physical POVM measurement $\{M_i\}$, it holds that the
expected SiV decreases:
\[\sum_i\langle\psi|M_i^\dagger N_A^2
M_i|\psi\rangle-\sum_i\frac{\langle\psi|M_i^\dagger N_A
M_i|\psi\rangle^2}{\langle\psi|M_i^\dagger M_i|\psi\rangle}\leq
\frac{1}{4}V(\psi)\] The inequality follows by using
Cauchy-Schwarz and the properties of the POVM-elements
($[M_i,N_A]=0,\sum_i M_i^\dagger M_i=\openone$). The SiV therefore
fulfills all requirements for a good entanglement measure, and we
expect it to completely characterize the particle number variation
in the asymptotic limit. This can be formalized in the following
central result of this paper:
\begin{theorem}
Consider $N$ copies of a state $|\phi\rangle$ with entropy of
entanglement $E(\phi)$ and SiV $V(\phi)$, then there exists an
asymptotically reversible conversion
\[
|\phi\rangle^N\leftrightarrow
\Big[|01\rangle_A|10\rangle_B+|10\rangle_A|01\rangle_B\Big]^{\otimes
NE(\phi)}\otimes \sum_n c_n|n\rangle_A|N-n\rangle_B,
\]
where the coefficients $c_n$ are Gaussian distributed with SiV
$NV(\phi)$, and $|n\rangle$ denotes the state
$|\underbrace{1\cdots 1}_{n}0\cdots 0\rangle$.
\end{theorem}

\emph{Proof:} In order to reduce the notational overhead, we will
outline the proof for a state
$|\phi\rangle=\sqrt{p_0}|0\rangle_A|1\rangle_B+\sqrt{p_1}|1\rangle_A|0\rangle_B$.
Given $N$ copies of the state $|\phi\rangle$, it holds that
\begin{eqnarray*}|\phi\rangle^{\otimes N}&=&\sum_n
\sqrt{c_n}|\psi_{N-n,n}\rangle\hspace{.5cm};\hspace{.5cm}
c_n=p_0^np_1^{N-n}\left(N\atop n\right)\\
|\psi_{N-n,n}\rangle&\propto&\sum_{{i_1i_2\cdots i_N=0}\atop{
\sum_ji_j=N-n}}^1|i_1\cdots i_N\rangle_A|1-i_1\cdots
1-i_N\rangle_B
\end{eqnarray*}
We have decomposed the state into components with constant local
particle number $|\psi_{N-n,n}\rangle$, and each of these is a
maximally entangled state with Schmidt number $\left(N\atop
n\right)$. In the limit of large $N$, the distribution of $c_n$
approaches a Gaussian with width $\propto\sqrt N$, and hence we
can restrict the sum over $n$ to the $\epsilon$-typical subspaces
$\mathcal{S}$ for which $\left(N\atop n\right)\geq
2^{N[H(p_0)-K\epsilon]}$ \cite{TC91}. Here $H(p)$ denotes the
Shannon entropy of the distribution $(p,1-p)$. Using the
previously derived majorization results, there exists an LOCC
protocol that can (coherently) decrease the Schmidt rank in all
$|\psi_{N-n,n}\rangle$ to the value $2^{N(H(p_0)-K\epsilon)}$
without changing their relative weight. By an appropriate
transformation of the local basis states, and by adding local
ancilla states to get the correct average particle number, this
new state can be rewritten as
\[
\Big[|01\rangle_A|10\rangle_B+|10\rangle_A|01\rangle_B\Big]^{\otimes
N[H(p)-K\epsilon]} \otimes
\sum_{n\in\mathcal{S}}c_n|N-n\rangle_A|n\rangle_B\] where $c_n$ is
a Gaussian distribution with variance $Np(1-p)=NV(\phi)/4$. Note
that the sum can again be extended to all $n$ with arbitrary
fidelity, and this completes the distillation direction of the
proof.

To prove the dilution step, it is enough to observe that, starting
from the state
\[
\Big[|01\rangle_A|10\rangle_B+|10\rangle_A|01\rangle_B\Big]^{\otimes
N[H(p)+K\epsilon]} \otimes
\sum_{n\in\mathcal{S}}c_n|N-n\rangle_A|n\rangle_B\] (note the term
$+K\epsilon$), the majorization step can be reversed and hence the
$\epsilon$-typical subspaces can be recovered. This completes the
proof for the case of \emph{qubits}. The proof for the general
case is completely analogous and will be presented elsewhere
\cite{SVC03b}.\qed

It follows that a state can be interconverted into a tensor
product of a state with only EoE and no SiV, and another part that
contains a negligible amount of EoE but all SiV. Indeed, the state
$|01\rangle_A|01\rangle_B+|10\rangle_A|10\rangle_B$ has constant
local particle number and therefore SSR impose no restrictions
whatsoever on the local operations that can be applied on this
part (see \cite{VC03}). On the other hand, the part with Gaussian
distributed coefficients contains all SiV and only a logarithmic
amount of entanglement ($\simeq \log(N)$), therefore solely
containing the SiV resources of the original state $|\phi\rangle$.
Note however that in the case of qubits, it is readily verified
that the EoE always exceeds the SiV, and an equivalent version of
the previous theorem would be:
\begin{eqnarray*}
|\psi\rangle^{\otimes N}&\leftrightarrow&
\Big[|01\rangle_A|10\rangle_B+|10\rangle_A|01\rangle_B\Big]^{\otimes
N[E(\psi)-V(\psi)]}\\
&&\hspace{1cm}\otimes
[|0\rangle_A|1\rangle_B+|1\rangle_A|0\rangle_B]^{\otimes NV(\psi)}
\end{eqnarray*}
As an example, this decomposition has been illustrated in Figure 1
for the state
$|\psi\rangle=\sqrt{1/6}|01\rangle+\sqrt{5/6}|10\rangle$.

\begin{figure}[t]
       \begin{center}
      \epsfig{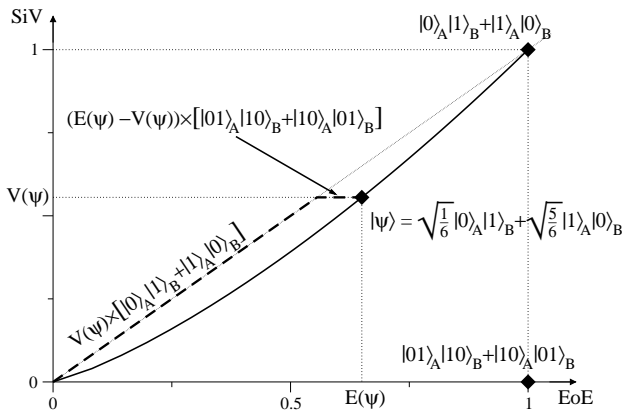}
    \end{center}
\caption{Diagram representing the two nonlocal resources
associated to a state
$|\psi\rangle=\sqrt{1/6}|01\rangle+\sqrt{5/6}|10\rangle$.}
       \label{fig1}
\end{figure}

The important insight of the given theorem is that, asymptotically
and in the presence of SSR, the Hilbert space decouples into a
tensor product of a Hilbert space equipped with a tensor product
structure and another one equipped with a direct sum structure,
the first one related to the EoE and the second one to the SiV:
$H_d^{\otimes n}\simeq H_d^{\otimes n-\log_dn}\otimes[\oplus_n
H_{d-1}]$.
The resource of EoE is very powerful and well
understood by now due to the concept of teleportation
\cite{BBC93}: it allows to completely overcome the constraints
imposed by LOCC as it allows to implement global operations
locally. As shown in \cite{VC03}, a similar result holds for the
SiV. Indeed, in the presence of SSR, it is impossible to
distinguish the Bell states
$|\phi^\pm\rangle=|01\rangle\pm|10\rangle$ by LOCC, which is the
basis for a perfect data hiding protocol; however, an extra state
$|01\rangle+|10\rangle$ allows to obtain (partial) information and
hence to the SiV can be used as a resource to overcome the
limitations induced by the SSR.

The crucial difference between the two resources however is their
associated Hilbert space structure: the dimension of the
respective Hilbert spaces grows exponential versus linear in the
number of qubits. A direct consequence is that e.g. the number of
bits that can be hidden securely \cite{VC03} in $N$ qubits scales
only as $\log_2(N)$. More fundamentally, we will show that this
implies that the resources of SiV needed to completely overcome
the locality constraints scale superlinearly in the number of
nonlocal particles under consideration: teleportation of
$N\rightarrow\infty$ particles is only possible using an amount of
maximally \emph{entangled} particles in the order of
$N^{1+\epsilon}$, such that the quotient of the Hilbert space of
the resources needed with that one of the state to be teleported
diverges. As already noticed in \cite{VC03}, this implies that in
the presence of SSR, having a bounded amount of
\emph{entanglement} and a classical channel is not equivalent
anymore to having a quantum channel: in the presence of SSR, it is
only possible to teleport one qubit with unit fidelity in the
limit of an unbounded amount of SiV.


Indeed, consider a resource state
$|\psi\rangle\propto\sum_{i=0}^M|i\rangle_{\bar{A}}|M-i\rangle_B$
which we would like to use to teleport Alice's part of a state
$|\phi\rangle=\sum_{j=0}^N\alpha_j|j\rangle_A|N-j\rangle_C$ with
$N$ particles shared between Alice and Charlie. Assuming $M\geq N$
and defining $\bar{n}=\min(n,N);\underline{n}=\max(0,n-M)$, the
total state is given by
\[\sum_{n=0}^{N+M}\sum_{j=\underline{n}}^{\bar{n}}\alpha_j|j,n-j\rangle_{A\bar{A}}|M-n+j\rangle_B|N-j\rangle_C.\]
Of course Alice can only act on the subspaces with constant
particle number $n$, and the best thing she can do is a
Bell-measurement in the local basis
$|\chi_k^{(n)}\rangle=1/\sqrt{\bar{n}-\underline{n}+1}\sum_{l=\underline{n}}^{\bar{n}}e^{i2\pi
lk/(\bar{n}-\underline{n}+1)}|l,n-l\rangle$ and communicate the
result to Bob who applies a unitary operation conditioned on her
result. Perfect teleportation has taken place iff
$\underline{n}=0$ and $\bar{n}=N$, and the probability that this
occurred is exactly given by $1-N/(M+1)$. It follows that perfect
teleportation is only possible when $M$ scales superlinearly with
$N$, proving the previously stated result. We conclude that both
EoE and SiV represent resources that allow for teleportation,
although the qualitative behavior is different.

The fact that SiV represents a resource independent of EoE can
also be appreciated by looking at the case of mixed states. As
shown in \cite{VC03}, there exist mixed states that are not
entangled in the usual sense but cannot be prepared locally due to
the SSR-constraints. In the light of SiV, this means that they are
nonlocal because SiV is needed to create them (but no EoE), and it
was furthermore shown that these mixed states can be used to do
things that are otherwise impossible. The present results allow to
turn these arguments into quantitative ones.

Therefore, it is natural to define the entanglement and the
variance of formation as
\begin{eqnarray*}
E_F^{\SSR}(\rho)&=&\min_{\{i,\psi_i\}}\sum_i p_iE(\psi_i)\\
V_F^{\SSR}(\rho)&=&\min_{\{p_i,\psi_i\}}\sum_i p_iV(\psi_i)
\end{eqnarray*}
where the ensemble $\{\psi_i\}$ has to be conform to the SSR. The
entanglement cost and variance cost \cite{HHT01} are then defined
as the regularized versions of it
($E_c^\SSR(\rho)=\lim_{N\rightarrow\infty}E_F^\SSR(\rho^{\otimes
N})/N$ ;
$V_c^\SSR(\rho)=\lim_{N\rightarrow\infty}V_F^\SSR(\rho^{\otimes
N})/N$). As an example, consider the state
\[\rho=\frac{1}{4}\left(|00\rangle\langle 00|+|11\rangle\langle
11|+(|01\rangle+|10\rangle)(\langle 01|+\langle 10|)\right).\]
This state is separable in the usual sense, but
$E_F(\rho)=1/2=V_F(\rho)$. To calculate its entanglement cost, we
need the following result: given a density operator that is
separable when no SSR apply, then its entanglement cost in the
presence of SSR is zero. This can readily be proven as follows:
suppose $\rho^{\otimes
N}=\sum_ip_i|\chi_i\rangle|\psi_i\rangle\langle\psi_i|\langle\chi_i|$
where the $\psi_i,\chi_i$ do not obey SSR. It holds that
$\rho^{\otimes N}=\sum_n P_n\rho^{\otimes N}P_n$ where $P_n$ is
the global projector on the (global) subspace of $n$ particles.
Acting with $P_n$ on a separable state, at most $\log_2(N)$ EoE
can be created, and therefore the entanglement cost will behave as
$\log_2(N)/N$, concluding the proof.  On the other hand, the
variance cost for the particular state $\rho$ can easily be shown
to be additive, as the subspaces of constant local particle number
are pure \cite{SVC03b}.

Analogously to the normal case of mixed states, one can readily
define quantities like entanglement of distillation \cite{BDS96b}
and variance of distillation. The recurrence schemes of
\cite{BDS96} can be adapted to the SSR-case, and it can be shown
that any qubit state with a vanishing $E_c^{\SSR}$ but
nonvanishing $V_F^\SSR$ can be converted into any other one of
this class, and any state with a non-vanishing $E_c^{\SSR}$ can be
purified to a perfect singlet $|01\rangle+|10\rangle$. A more
elaborate exposition of these results will be presented elsewhere
\cite{SVC03b}.

In conclusion, we have developed the theory of entanglement for
bipartite quantum states subject to SSR. We made the obvious
observation that, both in the case of LOCC and SSR, there exist
resources that let you overcome the constraints. We identified two
independent resources that asymptotically completely characterize
the amount of nonlocality present in a pure state, and showed that
these lead to the necessary and sufficient condition for
interconvertibility of quantum states under LOCC. One resource is
characterized by the familiar entropy of entanglement, while the
other one by the superselection induced variance. We discussed how
the resources allow to overcome the LOCC- respectively the
SSR-constraints by teleportation, although the SiV is
fundamentally different from the EoE because of the direct sum
structure of its associated Hilbert space. We concluded this paper
by stating some intriguing properties for the entanglement
properties of mixed bipartite states subject to SSR.

This work was supported in part by the E.C. (RESQ QUIPRODIS) and
the Kompetenznetzwerk ``Quanteninformationsverarbeitung'' der
Bayerischen Staats\-re\-gie\-rung.

\bibliographystyle{unsrt}

\end{document}